# Phase Diagram of FeSe Deposited by Electrochemical Technique with Different Temperature and Voltage

Aichi Yamashita[1, 2‡], Masashi Tanaka[1, 3], Hiroyuki Takeya[1], and Yoshihiko Takano[1, 2]

[1] WPI-MANA, National Institute for Materials Science, 1-2-1 Sengen, Tsukuba, Ibaraki, 305-0047, Japan

[2] Graduate School of Pure and Applied Sciences, University of Tsukuba, 1-1-4 Tennodai, Tsukuba, Ibaraki, 305-8577, Japan

[3] Graduate School of Engineering, Kyushu Institute of Technology, 1-1 Sensui-cho, Tobata-ku, Kitakyushu-shi, Fukuoka, 804-8550, Japan

**Abstract**

High-quality crystalline FeSe is successfully synthesized in 5 minutes by an electrochemical deposition technique via increasing the solution temperature. The synthesis of FeSe have been controlled by the applied voltage, synthesis time and pH value. The obtained samples were summarized as a phase diagram at each temperature and applied voltage. The best synthesis temperature to produce high-quality crystalline samples was found to be 70°C.

Iron-based superconductors have potential application for use in under high magnetic fields because of their high superconducting transition temperature ($T_c$), upper critical field ($H_{c2}$), and irreversibility field ($H_{irr}$).[1,2] FeSe has the simplest crystal structure among the iron-based superconductors [3] and it is expected to be a candidate material for superconducting wires and tapes. The $T_c$ of FeSe increases from 8 K to 37 K by applying high pressure [4-7] or 30-46 K by the intercalation of alkali or alkaline earth elements. [8-10]

As a new way to synthesize FeSe superconductors, we have developed the electrochemical deposition technique to fabricate the superconducting FeSe films.[11-13] The electrochemical deposition is known as an easy and inexpensive method to fabricate films on a variety of substrates and over large areas, leading to the low cost fabrication of superconducting FeSe coated conductors by the reel to reel method. In this study, we have constructed a phase diagram clarifying the optimal synthesis condition as a function of temperature, applied voltage for FeSe grown by electrochemical deposition. We have succeeded in synthesizing the high-quality crystalline superconducting FeSe at a temperature of 70°C in a relatively short time.

Electrochemical deposition was carried out by a three-electrode method with Indium Tin Oxide (ITO) as the working electrode, the platinum sheet as the counter electrode and the Ag/AgCl as the reference electrode. An aqueous solution was prepared by dissolving 0.03 M of $FeCl_2 \cdot 4H_2O$, 0.015 M of $SeO_2$ and 0.1 M of $Na_2SO_4$ in distilled water. The pH of the solution was adjusted by adding $H_2SO_4$. All experiments were performed with a pH value of 2.1 as was previously reported.[12,13] The optimal synthesis conditions were investigated by changing the applied voltage against various temperatures (20, 50, 70, 90°C). The reaction time was fixed at 5 minutes in all experiments. Powder X-ray diffraction (XRD) measurements were performed using Mini Flex 600 (Rigaku) with Cu-K$\alpha$ radiation. The temperature dependence of magnetic susceptibility was measured using a SQUID magnetometer (Quantum Design MPMS) down to 2.0 K under a field of 10 Oe.

At this current stage, the role which temperature plays in the formation of FeSe films using an electrochemical deposition method is unclear and therefore we sought to perform a synthesis study focusing on an increase in solution temperature and the effect which it has on the formation of high-quality crystalline FeSe. We have investigated the XRD patterns as a function of the applied voltage and each temperature. The overall results of phase diagram are summarized in fig.1 (a). By varying the voltage and temperature the results largely can be categorized into three regions. The Se phase exists at a lower bias voltage with temperature above 70°C. A mixed phase of FeSe and Se appears when increasing the bias voltage. FeSe single phase region was obtained from the applying voltages -0.7 to -1.1 V, -1.3 to -1.7 V, -1.3 to -2.5 V and at -2.7 V with temperatures of 20, 50, 70 and 90°C, respectively. We found that the region moved to higher bias voltage with increase of temperature. The high-quality crystalline FeSe films were obtained in 5 minutes at temperature 70°C. This fabrication time is much shorter than that of 1 hour at 20°C in our previous reports.[11-13]

Figure 1 (b) shows the typical XRD patterns in the three regions at a temperature of 70°C. Only Se was detected at -0.9 V and FeSe started to grow with increase of applied voltage above -1.3 V and single phase appeared above -1.5 V. The sharpest diffraction peaks of FeSe were obtained at -1.7 and -1.9 V. This behavior implies that the fabrication of FeSe and Se can be controlled by changing the applied voltage. Figure 1 (c) shows the XRD patterns in the FeSe single phase region using the conditions described by the red line in fig.1 (a). Among them, tetragonal FeSe deposited at -1.7 V at a temperature of 70°C showed the sharpest peaks, indicating that it is the best condition for the high-quality crystalline FeSe. We also observed a superconducting transition for this sample at around 8 K in the magnetization measurement.

In conclusion, we have succeeded in synthesizing the single phase tetragonal FeSe with sharp diffraction peaks just in 5 minutes at temperature 70°C. We found that the applied voltage for single phase of FeSe moved to higher bias voltage with increase of temperature. The crystallinity of tetragonal FeSe films synthesized at 70°C found to be the best among that of 20, 50 and 90°C. Further improvement of the synthesis conditions leads to obtain the superconducting FeSe film which shows the zero resistivity. This electrochemical deposition technique will be the promising way to fabricate the superconducting FeSe film, wire and tape in a short time.

**Acknowledgments**
This work was partly supported by JST CREST, Japan, and JSPS KAKENHI Grant Number JP16J05432.

‡ Corresponding author: Aichi Yamashita

E-mail: YAMASHITA.Aichi@nims.go.jp

Postal address: National Institute for Materials Science, 1-2-1 Sengen, Tsukuba, Ibaraki 305-0047, Japan

Tel.: (+81)29-851-3354 ext. 2976

**References**

[1] S. Khim, J. W. Kim, E. S. Choi, Y. Bang, M. Nohora, H. Takagi, and K. H. Kim, Phys. Rev. B 81, 184511 (2010).

[2] J. D. Weiss, C. Tarantini, J. Jiang, F. Kametani, A. A. Polyanskii, D. C. Larbalestier & E. E. Hellstrom, Nat. Mater. 11, 682 (2012).

[3] F. C. Hsu, J. Y. Luo, K. W. Yeh, T. K. Chen, T. W. Huang, P. M. Wu, Y. C. Lee, Y. L. Huang, Y. Y. Chu, D. C. Yan and M. K. Wu, PNAS, 105 (2008), p.14262

[4] Y. Mizuguchi, F. Tomioka, S. Tsuda, T. Yamaguchi, and Y. Takano, Appl. Phys. Lett. 93, 152505 (2008).

[5] S. Margadonna, Y. Takabayashi, Y. Ohishi, Y. Mizuguchi, Y. Takano, T. Kagayama, T. Nakagawa, M. Takata, and K. Prassides, Phys. Rev. B 80, 064506 (2009).

[6] S. Medvedev, T. M. McQueen, I. A. Troyan, T. Palasyuk, M. I. Eremets, R. J. Cava,

S. Naghavi, F. Casper, V. Ksenofontov, G. Wortmann, and C. Felser, Nat. Mater. 8, 630 (2009).

[7] S. Masaki, H. Kotegawa, Y. Hara, H. Tou, K. Murata, Y. Mizuguchi, and Y. Takano, J. Phys. Soc. Jpn. 78, 063704 (2009).

[8] Y. Mizuguchi, H. Takeya, Y. Kawasaki, T. Ozaki, S. Tsuda, T. Yamaguchi, and Y. Takano, Appl. Phys. Lett. 98, 042511 (2011).

[9] T. P. Ying, X. L. Chen, G. Wang, S. F. Jin, T. T. Zhou, X. F. Lai, H. Zhang, and W. Y. Wang, Sci. Rep. 2, 426 (2012).

[10] M. Tanaka, Y. Yanagisawa, S. J. Denholme, M. Fujioka, S. Funahashi, Y. Matsushita, N. Ishizawa, T. Yamaguchi, H. Takeya and Y. Takano: J. Phys. Soc. JPN. 85(2016) 044710.

[11] S. Demura, T. Ozaki, H. Okazaki, Y. Mizuguchi, Y. Kawasaki, K. Deguchi, T. Watanabe, H. Hara, T. Yamaguchi, H. Takeya, and Y. Takano, J. Phys. Soc. Jpn. 81, 043702 (2012).

[12] S. Demura, H. Okazaki, T. Ozaki, H. Hara, Y. Kawasaki, K. Deguchi, T. Watanabe, S. J. Denholme, Y. Mizuguchi, T. Yamaguchi, H. Takeya, and Y. Takano, Solid State Commun. 154, 40 (2013).

[13] S. Demura, M. Tanaka, A. Yamashita, S. J. Denholme, H. Okazaki, M. Fujioka, T.

Yamaguchi, H. Takeya, K. Iida, B. Holzapfel, H. Sakata and Y. Takano: J. Phys. Soc. Jpn. 85, 015001 (2016)

**Figure caption**

Fig. 1 (a) Phase diagram of the obtained samples at various temperatures and applied voltage. (b) XRD patterns of tetragonal FeSe deposited on ITO substrate at applied voltage -0.9, -1.3 and -1.7 V at 70°C. All the indices indicate tetragonal FeSe. ITO substrate and Se are shown as + and asterisk, respectively. (c) XRD patterns of the highest crystallinity samples at each solution temperature 20, 50, 70 and 90°C. The sample synthesized at -1.7 V and 70°C showed the best crystallinity in the conditions.

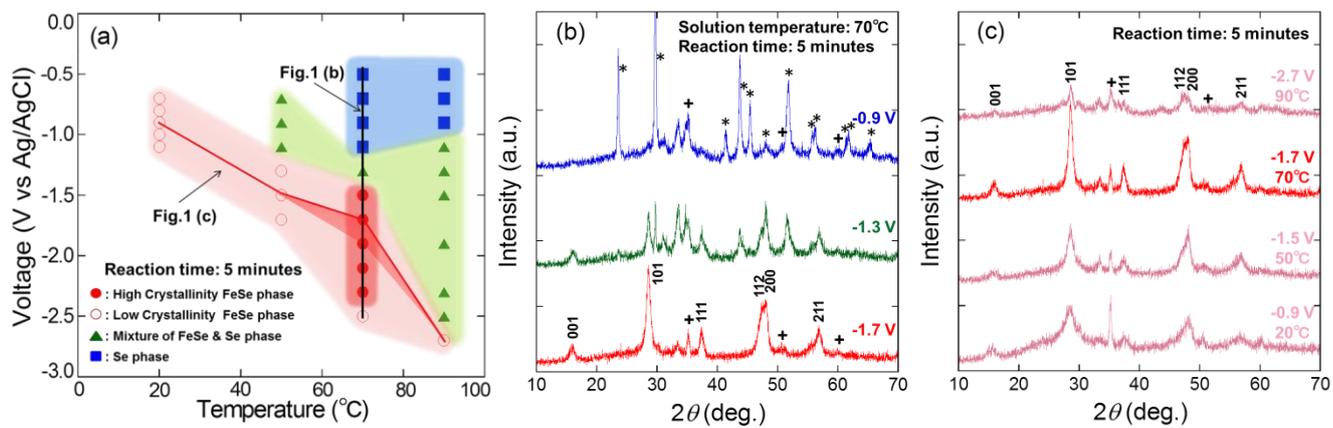

Fig.1